\documentclass[twocolumn,prl,amsmath,amssymb,floatfix,showpacs]{revtex4}
\usepackage{graphicx}
\usepackage{bm}
\begin{document}
\title{High T$_c$ superconductivity in superhard diamond-like BC$_{5}$}

\author{Matteo Calandra}
\author{Francesco Mauri}
\affiliation{CNRS and 
Institut de Min\'eralogie et de Physique des Milieux condens\'es, 
case 115, 4 place Jussieu, 75252, Paris cedex 05, France}
\date{\today}

\begin{abstract}
Using density functional theory calculations we show that the recently synthesized
\cite{LeGodec} superhard diamond-like BC$_5$ is superconducting with a critical 
temperature of the same order than that of MgB$_2$.
The average electron-phonon coupling is $\lambda=0.89$, 
the phonon-frequency logarithmic-average is $\langle\omega \rangle_{\log}=67.4 $ meV and
the isotope coefficients are $\alpha(C)=0.3$ and $\alpha(B)=0.2$.
In BC$_5$, superconductivity is mostly sustained by vibrations of
the B atom and its C neighbors.
\end{abstract}
\maketitle


Diamond is the hardest material available in nature,
a very good thermal conductor and a large gap insulator. 
Heavily Boron-doped diamond (B$_{\delta}$C$_{1-\delta}$) 
is superconducting at Boron content  $\delta=0.028$ 
 \cite{Ekimov}.
In thin films, T$_c$ increases with $\delta$ up to 7K 
\cite{Bustarret}. These critical temperatures
are lower than those of other superconducting
carbon materials such as nanotubes \cite{Tang} (T$_c=15 $K),
intercalated graphite compounds 
\cite{Weller, Genevieve} (T$_c=11.4$ K in CaC$_6$) and 
alkali-doped fullerides \cite{Hebard,Tanigaki} 
(T$_c=33$ K in RbCs$_2$C$_{60}$). Thus, hole-doping of 
Diamond doesn't look very promising for high T$_c$ 
superconductivity.
However, very recently, superhard nanocrystalline aggregates of BC$_{5}$, 
corresponding to B-content of $\delta=0.166$
 were synthesized \cite{LeGodec}. Besides being a superhard
material, harder than c-BN, BC$_5$ is interesting since the 
introduction of such a massive number of carriers in diamond can
induce strong scattering between holes and hard phonon modes.

At low B-content  ($\delta \approx 0.02-0.06$), 
Density Functional Theory (DFT) calculations based on a supercell
approach \cite{Blase, Xiang, GiustinoPRL} have shown that superconductivity
is phonon mediated with holes scattering to 
high energy optical phonon modes. The contributions of the 
acoustic modes is negligible. 
Most surprisingly, B phonon states play a
major role in the coupling to optical modes, despite the low B concentration. 
Moreover a L\"owdin population analysis shows that the electronic states
at $\epsilon_f$ are mainly of B-character \cite{Blase}.
This is contrary to what happen in rigid band doping \cite{Lee} or in 
a Virtual Crystal Approach \cite{Boeri, Ma, GiustinoPRL}.

The coupling to hard optical phonons generates a large electron-phonon matrix element.
However, at low doping, the average electron-phonon coupling ${\lambda}$
and T$_c$ are small because of the low number of carriers. 
Indeed the increase of T$_c$ with doping observed in thin Films \cite{Bustarret}
is mainly to due the increase in the 
density of states (DOS) \cite{Xiang} at the Fermi level. 

These findings are valid at dopings of the order of some percents, but are 
questionable at dopings as large as those of BC$_{5}$. First of all it is not
clear at large doping how much the number of
carriers can increase, since a rigid band picture fails and the results
depends on the fine details of the DOS. 
Furthermore, if many electronic states are allowed to couple to
phonons generating a considerable electron-phonon coupling, the system
can be driven to a charge density wave insulator.
Thus the occurrence of superconductivity in BC$_{5}$ cannot
be inferred from low doping results.

In this work, using density functional theory calculations, 
we demonstrate that BC$_{5}$ has a critical temperature of the same
order of that of MgB$_2$ \cite{Nagamatsu}.

Cubic BC$_5$ occurs as nanocrystalline aggregates \cite{LeGodec}. 
The crystal structure has, on average, cubic symmetry with volume per atom
of $6.00 \AA^3$. This value is $6\%$  larger than
the diamond volume per atom, $ 5.67 \AA^3 $.
Due to their similar atomic numbers, diffraction cannot distinguish between B
and C atoms and consequently the position of B in the cell is not determined.
In order to resolve the BC$_{5}$ crystal structure we start with the 
6-atoms hexagonal supercell of the 2 atoms diamond cell.
The hexagonal supercell is compatible with a cubic symmetry for
$c/a=\sqrt{6} \approx 2.45 $. The diamond theoretical equilibrium structure
has $a=2.529 \AA$.
Then we replace a C atom with a B atom and we perform volume and force
optimization \cite{comp_details}. 
The results are reported in tab.  \ref{tab:structure}.
We find  $a=2.55\AA$ and $c/a=2.50$. The theoretical volume per atom   
in BC$_5$ is $5.2\%$ larger
than in diamond, in very good
agreement with experiments. The $c/a$ found is only slightly larger than the
ideal one for cubic symmetry so that the BC$_5$ cell can be seen as a small 
elongation of the cubic cell along the cubic $(1 1 1)$ axis.
The most stable structure is non-magnetic.
Then we consider a $12$ atoms cell obtained replicating the hexagonal cell
along $c$  and consider all the possible
positions for two B atoms in this supercell. We found the configuration 
corresponding to the 6 atoms hexagonal unit cell 
to be the most stable by $5$ mRyd/atom.

\begin{table}[hbt]
\begin{ruledtabular}
\begin{tabular}{cccc}
Atom type &   X   &   Y   &   Z   \\ \hline
    B     &   0.0    &  0.0    & 0.0070  \\
    C     &   0.0    &  0.0    & 0.2592  \\
    C     &   1/3    &  -1/3   & 0.3386  \\
    C     &   1/3    &  -1/3   & 0.5816  \\
    C     &   2/3    &  -2/3   & 0.6646  \\
    C     &   2/3    &  -2/3   & 0.8999  \\
\end{tabular}
\end{ruledtabular}
\caption{Reduced coordinates of the theoretically devised atomic structure
of BC$_{5}$. The crystal symmetry is hexagonal with $a=2.55\AA$ and
$c/a=2.50$.}
\label{tab:structure}
\end{table}

\begin{figure}[h]
\hbox{0.1}\includegraphics[width=8.0cm]{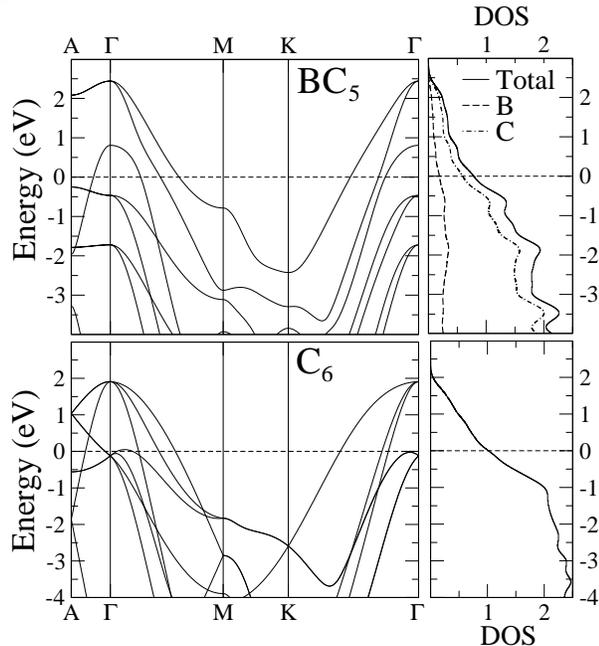}
\caption{BC$_{5}$ and diamond (C$_6$) band structure and DOS in the hexagonal
Brillouin zone using the BC$_5$ lattice parameters.
The DOS is in units of states/eV/(6 atoms cell). In C$_6$ the Fermi level
corresponds to rigid band doping of diamond.} 
\label{fig:Bands_and_dos}
\end{figure}

The BC$_{5}$ and diamond (C$_6$) electronic structures 
along high energy directions in the 
hexagonal Brillouin zone are plotted plotted in Fig. 
\ref{fig:Bands_and_dos}. For C$_6$ we used the BC$_5$ lattice
parameters and the Fermi level refers to a rigid band doping
of diamond of 1 hole per C$_6$ unit. As it can be seen,
the electronic structure of BC$_5$
cannot be interpreted as rigid band doping of diamond.

In BC$_5$, the Fermi level $\epsilon_f$ 
is shifted of 2.44 eV respect to the top of the valence band, to
be compared with a shift $\approx 0.8$ eV at $\delta=0.0278$ \cite{Xiang}.
The DOS at $\epsilon_f$ is $N(0)=0.78$ states/eV/(6 atoms cell), 
$2.1$ times larger than its value at $\delta=0.0278$ \cite{Xiang}, meaning
that the number of carriers induced by B-doping grows continuously without saturating
even for such large $\delta$. L\"owdin population analysis  
demonstrates that at $\epsilon_f$ the total B DOS is 
$N_B(0)=0.19$, states/eV/(6 atoms cell), slightly larger that  $N(0)/6$. Thus,
contrary to what happens at low doping \cite{Blase}, 
the number of carriers in B electronic states is comparable to the B-content.

\begin{figure*}[t]
\includegraphics[width=9.0cm]{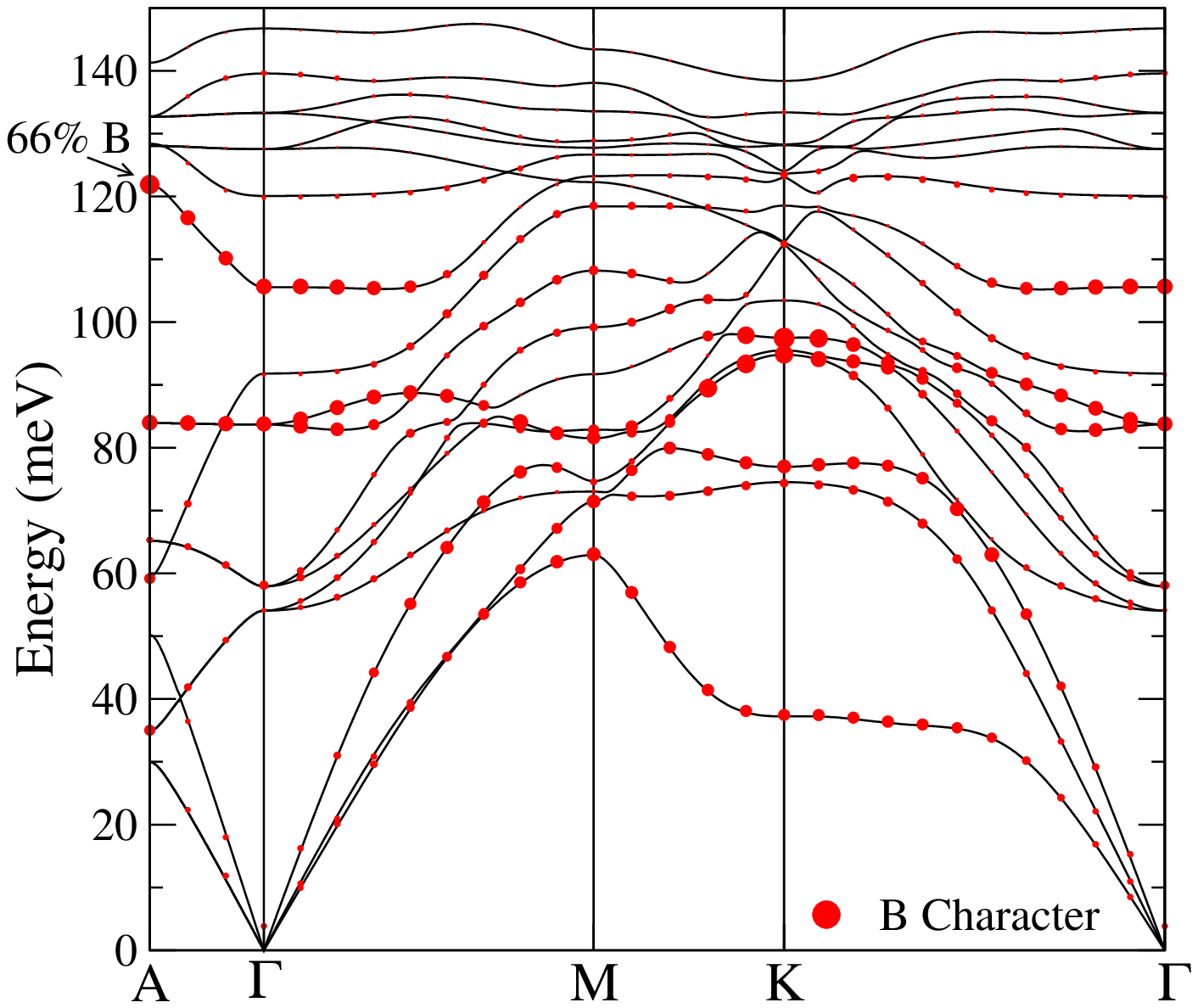}%
\includegraphics[width=9.0cm]{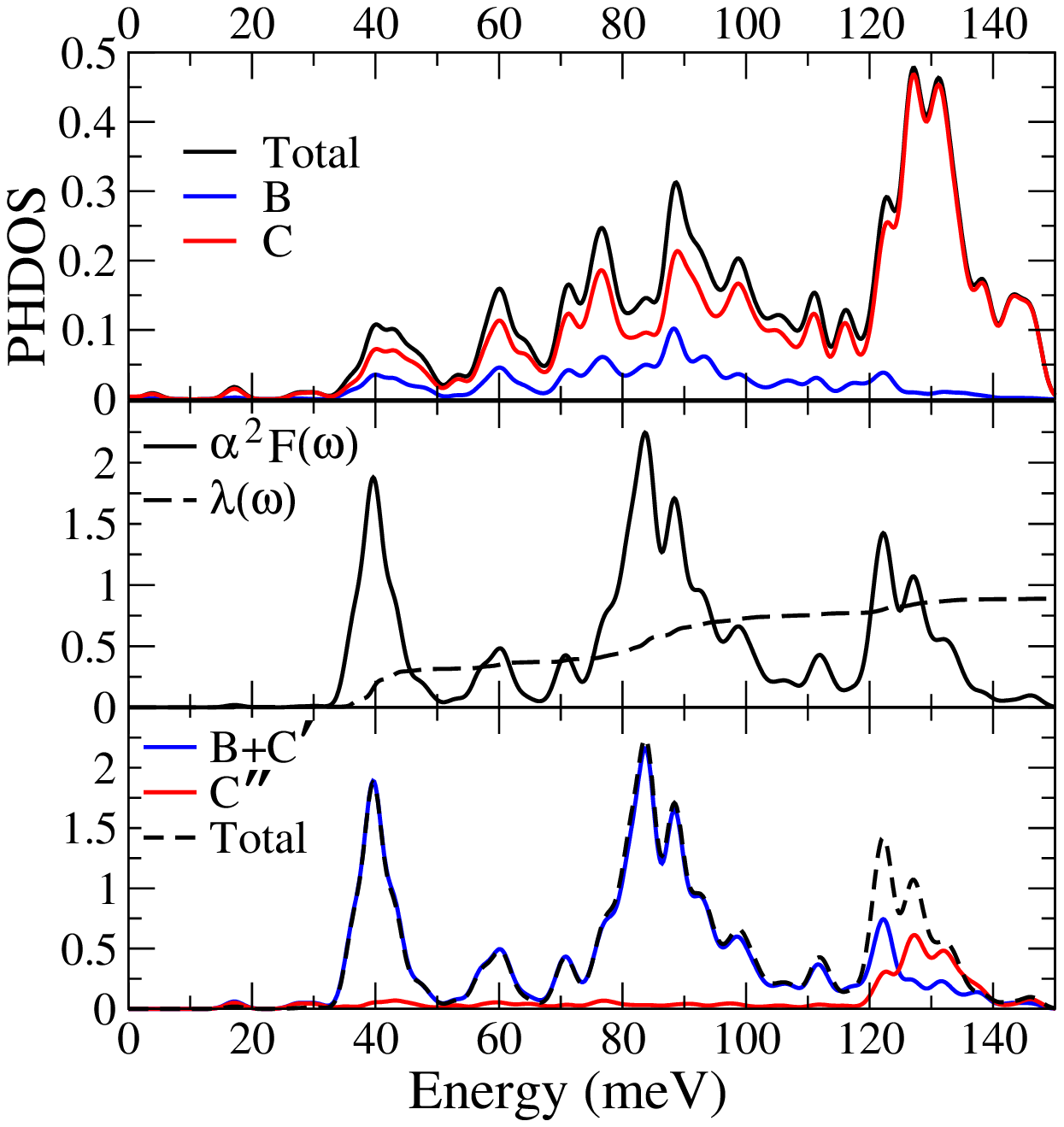}%
\caption{(Color). (Left) BC$_5$ phonon dispersion in the hexagonal
Brillouin zone. The radius of the red dots is proportional to the B content
of the branch. (Right) BC$_5$ 
Phonon density of states projected over B and C states, Eliashberg function 
$\alpha^2F(\omega)$ and
integrated electron-phonon coupling $\lambda(\omega)$. The Eliashberg function has
been decomposed three different contributions, due to vibrations of B and
its C neighbors (BC'), and vibrations of far neighbors (C").}
\label{fig:branchie_a2F}
\end{figure*}

The BC$_{5}$ phonon dispersion \cite{comp_details} 
is shown in Fig.~\ref{fig:branchie_a2F}.
No dynamical instabilities are seen in the phonon spectrum
meaning that the crystal structure obtained with geometrical optimization is
dynamically stable. No charge density wave instabilities occur
in the system.
Decomposition of phonon vibrations (Fig. \ref{fig:branchie_a2F} )  
into atomic components 
shows that, despite B being lighter than C, the harder phonon modes 
(120-150 meV) are due to C-vibrations.
At lower energies
($< 100$ meV) the B component in the phonon density of states (PHDOS)
is from $1/3$ to $1/2$ of the C-component.

The superconducting properties can be understood calculating the
electron-phonon coupling $\lambda_{{\bf q}\nu}$ for a phonon mode
$\nu$ with momentum ${\bf q}$ and phonon frequency $\omega_{\bf q \nu}$, namely:

\begin{equation}\label{eq:elph}
\lambda_{{\bf q}\nu} = \frac{4}{\omega_{{\bf q}\nu}N(0) N_{k}} \sum_{{\bf k},n,m} 
|g_{{\bf k}n,{\bf k+q}m}^{\nu}|^2 \delta(\epsilon_{{\bf k}n}) \delta(\epsilon_{{\bf k+q}m})
\end{equation}
where the sum is performed over a $N_k=40^3$ k-point mesh in the Brillouin Zone.
The matrix element is
$g_{{\bf k}n,{\bf k+q}m}^{\nu}= \langle {\bf k}n|\delta V/\delta e_{{\bf q}\nu} |{\bf k+q} m\rangle /\sqrt{2 \omega_{{\bf q}\nu}}$,
where 
 $V$ is the Kohn-Sham potential and 
$e_{{\bf q}\nu}=\sum_{A \alpha}M_A\sqrt{2 \omega_{{\bf q}\nu}}
\epsilon_{A\alpha}^{{\bf q}\nu} u_{{\bf q}A\alpha}$. 
 $u_{{\bf q}A\alpha}$ is the
Fourier transform of the $\alpha$ component of the
phonon displacement of the atom $A$ in the unit cell,
$M_A$ is the mass of atom $A$ and $\epsilon_{A\alpha}^{{\bf q}\nu}$
are $A\alpha$  components of ${\bf q}\nu$ phonon eigenvector 
normalized on the unit cell.

The average electron-phonon coupling is  
$\lambda=\sum_{{\bf q}\nu} \lambda_{{\bf q}\nu}/N_q\approx0.89$.
Thus BC$_5$ is a moderate coupling superconductor with $\lambda$
comparable to MgB$_2$.
To address the role of the different vibrations in determining
the electron-phonon coupling we decompose the electron-phonon 
coupling \cite{Calandra2005} into selected atomic vibrations, using 
the relation:
\begin{equation}\label{eq:trlambda}
\lambda=\sum_{i, j}\Lambda_{i, j}=
\sum_{A,B}\left[\sum_{\alpha, \beta}\frac{1}{N_q}\sum_{\bf q}
 [{\bf G}_{\bf q}]_{A\alpha,B\beta} [{\bf C_q}^{-1}]_{B\beta,A\alpha}\right]
\end{equation}
where $A,\alpha$ ($B\beta$) indicates the displacement of the 
$A^{\rm th} (B^{th})$ 
atom in $\alpha (\beta)$ Cartesian direction,
$[{\bf G_q}]_{A\alpha,B\beta}=\sum_{{\bf k},n,m}4 {\tilde
g}_{A\alpha}^{*}{\tilde g}_{B\beta} \delta(\epsilon_{{\bf k}n})
\delta(\epsilon_{{\bf k+q}m})/[N(0) N_{k}]$, and ${\tilde
g}_{A\alpha}=\langle {\bf k}n|\delta V/\delta u_{{\bf q} A\alpha}
|{\bf k+q} m\rangle /\sqrt{2}$.
The ${\bf C_q}$ matrix is the Fourier
transform of the force constant matrix (the derivative of the forces
with respect to the atomic displacements).
The decomposition \cite{footnote} leads to the following matrix 
($\lambda=\sum_{i,j}\Lambda_{i,j}$):
\begin{equation}
  \bm{\Lambda}\,=
  \begin{matrix}
    &
    \begin{matrix}
     \,\, {\rm B} & \,\,\,\,\,\, {\rm C'} & \,\,\,\,\,\,{\rm  C''}  \\
    \end{matrix} \\
    \begin{matrix}
       {\rm B}            \\
       {\rm C'} \\
       {\rm C''}     \\
    \end{matrix} &
    \begin{pmatrix}
        0.50  &  -0.18  &  0.04   \\
       -0.18  &   0.69  & -0.07    \\
        0.04  &   -0.07 &  0.12    \\
    \end{pmatrix}\\
    \begin{matrix}
       &   &  \\
    \end{matrix} \\
  \end{matrix}
  \label{eq:Lambda}
\end{equation}
where C$'$ stands for the first two C nearest neighbors of the B atom,
and C$''$ stands for the other three C atoms in the cell.
The dominant contribution to the electron phonon-coupling comes from 
the vibrations of the B atom and its C neighbors. 

The Eliashberg function
\begin{equation}
\alpha^2F(\omega)=\frac{1}{2 N_q}\sum_{{\bf q}\nu} \lambda_{{\bf q}\nu} \omega_{{\bf q}\nu} \delta(\omega-\omega_{{\bf q}\nu} )
\end{equation}
and the integral $\lambda(\omega)=2 \int_{0}^{\omega} d\omega^{\prime} 
\alpha^2F(\omega^{\prime})/\omega^{\prime}$ 
are shown in Fig. \ref{fig:branchie_a2F} . $\alpha^2F(\omega)$ is composed by
three main peaks. This is different from the low
doping case \cite{GiustinoPRL} since in BC$_5$
the high energy modes (80-90 meV and 130 meV) are softer due to higher doping
and the low energy modes ($\approx 40$ meV) are significantly
coupled (at low doping the coupling is very weak).

To understand the role of different vibrations in 
determining the total electron-phonon coupling, we decompose $\alpha^2F$
as:
\begin{eqnarray}
\alpha^{2}F(\omega)&=&\sum_{A,B} \alpha^{2}_{A,B}F(\omega) = \nonumber \\
&=&\sum_{A\alpha,B\beta} 
\left[
\frac{1}{N_{\bf q}}\sum_{\bf q}
\frac{ [{\rm G}_{\bf q}]_{A\alpha,B\beta}[{\rm L}_{\bf q}]_{B\beta,A\alpha}}{2 \sqrt{M_A M_B}}
\right]
\end{eqnarray}
where 
$[{\rm L}_{\bf q}]_{B\beta,A\alpha}=\sum_{\rho} 
\epsilon_{{\bf q}\rho}^{A\alpha}
\frac{\delta(\omega-\omega_{{\bf q}\rho})}{\omega_{{\bf q} \rho}}
(\epsilon_{{\bf q}\rho}^{B\beta})^{*}$. 
Then we plot the decomposition of $\alpha^2F(\omega)$ into 
vibrations of B and its first two C nearest neighbors (BC') and vibrations 
of the other three C atoms in the cell (C'').
 As it can be seen in Fig. \ref{fig:branchie_a2F}, the contribution  
of the highest energy mode (115-160 meV),
essentially due to vibrations of the C-C bonds, to the electron-phonon coupling is 
$\approx 0.13$. This is half of the contribution of the high energy 
structure to $\lambda$ in doped-diamond at $\delta=0.018$
(from the inset of fig. 3 in ref. \cite{GiustinoPRL} one
can infer that the contribution of the lower energy modes is $\approx 0.09$),
indicating that in BC$_{5}$ substantial weight is shifted at
low energy. At lower energies, the other two main peaks in $\alpha^2F(\omega)$
are entirely due to B and its C neighbors. 
Further decomposition of $\alpha^2F(\omega)$ onto vibration of B and C atoms
(not shown) indicates that the two low energy peaks are due to phonon modes
with concerted movements of BC atoms.
Thus, despite the lowest B concentration, the contribution of vibrations 
associated to the B atom and its neighbors are the main responsible for
superconductivity.

As demonstrated, BC$_{5}$ has two times larger number of carriers
than doped-diamond at $\delta=0.0278$ and the phonons coupled to electrons are
less energetic. Since $\lambda_{\bf q \nu}$ is proportional $N(0)/\omega_{\bf q \nu}^2$,
both these effects cooperates in increasing $\lambda$. 

The critical superconducting temperature is estimated using the McMillan 
formula \cite{mcmillan},
\begin{equation}
T_c = \frac{\langle \omega \rangle_{\log}}{1.2}\, \exp \left[{ - \frac{1.04 (1+\lambda)}{\lambda-\mu^* (1+0.62\lambda)}}\right]
\label{eq:mcmillan}
\end{equation}
where $\mu^*$ is the screened Coulomb pseudopotential and 
$\langle\omega\rangle_{\log} = \exp[\frac{2}{\lambda}\int_{0}^{+\infty} 
\alpha^2F(\omega)\log(\omega)/\omega\,d\omega ]$ is
the phonon frequencies logarithmic average. We obtain 
$\langle\omega\rangle_{\log}=67.4$ meV, to be compared with $105$ meV
in doped diamond at $\delta=0.0278$\cite{Xiang} and $62$ meV
in MgB$_2$\cite{Kong}. The reduction in $\langle\omega\rangle_{\log}$ as
compared to hole doped diamond is due to (i) activation of coupling to low
energy modes which are very weakly coupled in B-doped diamond and 
(ii) softening of the high energy modes due to the larger doping.

Using the same value of 
$\mu^{*}=0.1$ necessary for hole-doped diamond in order to 
obtain T$_c = 4$K \cite{Xiang} we
obtain T$_c=45$ K, which is larger than the experimentally measured
39 K in MgB$_2$\cite{Nagamatsu} and puts BC$_{5}$ in the class of
high T$_c$ superconductors. As in MgB$_2$, the T$_c$ could be further increased
by multi-band effects\cite{Multiband}

The mutual relevance of B and C phonon modes in sustaining 
superconductivity in BC$_5$ can be addressed measuring the isotope
effect coefficients for a given atomic specie $X={\rm B},{\rm C}$, namely
$\alpha(X)=-(d\log T_c/d M_X)$. We obtain $\alpha(C)=0.3$ and 
$\alpha(B)=0.2$, confirming the important role of B-phonon modes.

A question arise whether the actual samples of BC$_5$ are able to sustain
superconductivity, due to the reduced size of the grains (diameter 10-15 nm)
\cite{LeGodec}.
The observation of Bulk superconductivity is possible only if
if the coherence length $\xi_0 \sim \hbar v_F/\Delta$ 
is at least comparable to the size of the grains.
In B-doped diamond \cite{Sacepe}, $\xi_0=240 $ nm. Assuming parabolic
bands and the same
$\Delta(0)/k_b T_c=1.78$ as in Boron-doped diamond, 
$v_f$ is 16 times larger and  $\xi_0 \approx 400$ nm.
Thus it is necessary to grow larger samples to 
observe bulk superconductivity in BC$_5$, possibly by longer
synthesis or by the use of catalysts to speed up the reaction.

High energy phonon modes can lead to large T$_c$s  
even with moderate electron-phonon scattering ($\lambda\sim 0.8-1.0$),
as in MgB$_2$. Interestingly, the quest for such a kind of 
high-temperature phonon-mediated superconductors 
coincides with that of metallic superhard materials.
Superhard materials 
have large elastic constants, requiring short and
strong chemical bonds, typically found in light-element compounds.
These conditions result in energetic phonon modes.  
Thus high T$_c$ superconductivity could be realized in superhard
materials if doping
is large enough to sustain a moderate electron-phonon coupling $\lambda$.

In this work we have shown that this
happens in the recently-synthesized \cite{LeGodec} superhard BC$_{5}$ 
which is predicted to be metallic and superconducting
with a T$_c=45$ K, the largest T$_c$ ever for
a phonon-mediated superconductor. 

We acknowledge discussions with Yann le Godec and M. d'Astuto and A. Gauzzi.
Calculations were performed at the IDRIS supercomputing center (project 081202).


\begin{thebibliography}{99}
\bibitem{LeGodec}  V.L. Solozhenko, D. Andrault, O.O. Kurakevych, Y. Le Godec, 
M. Mezouar, submitted (see additional material sent to the Editor).

\bibitem{Ekimov} E. A. Ekimov {\it et al.}, Nature {\bf 428}, 542 (2004)

\bibitem{Bustarret} E. Bustarret {\it et al.}, 
Phys. Rev. Lett. {\bf 93}, 237005 (2004)

\bibitem{Tang} Z. K. Tang {\it et al.}, 
Science {\bf 292}, 2462 (2001)

\bibitem{Weller} T. E. Weller {\it et al.}, Nature Phys. {\bf 1}, 39 (2005) 

\bibitem{Genevieve} N. Emery {\it et al.}, Phys. Rev. Lett. {\bf 95}, 087003 (2005)

\bibitem{Hebard} A. F. Hebard {\it et al.}, 
Nature {\bf 350}, 600 (1991).

\bibitem{Tanigaki} K. Tanigaki {\it et al.}, Nature {\bf 352}, 222, (1991).

\bibitem{Blase}
X. Blase, Ch. Adessi, and D. Conn\'etable, 
Phys. Rev. Lett. {\bf 93}, 237004 (2004).

\bibitem{Xiang}
H. J. Xiang {\it et al.}, 
Phys. Rev. B {\bf 70}, 212504 (2004). 

\bibitem{GiustinoPRL} F. Giustino {\it et al.},
Phys. Rev. Lett. {\bf 98}, 047005 (2007)

\bibitem{Lee} K. W. Lee and W. E. Pickett, 
Phys. Rev. Lett. {\bf 93}, 237003 (2004).

\bibitem{Boeri} L. Boeri, J. Kortus, and O. K. Andersen, 
Phys. Rev. Lett. {\bf 93},237002 (2004)

\bibitem{Ma} Y. Ma {\it et al.}, 
Phys. Rev. B {\bf 72}, 014306 (2005) 

\bibitem{Nagamatsu} J. Nagamatsu {\it et al.},
Nature {\bf 410}, 63 (2001).

\bibitem{comp_details}
Density functional theory calculations \cite{PWSCF} were 
performed in the generalized gradient approximation \cite{PBE} using norm-conserving 
pseudopotentials and a 65 Rydberg cutoff. 
The electronic integration is performed using a 
$14\times 14 \times 14$ k-point mesh and a $60$ mRyd Hermite-Gaussian smearing
of order 1. The phonon dispersion is obtained by Fourier interpolation of the 
Dynamical matrices calculated over a $4\times 4\times 4$ k-point mesh.

\bibitem{PWSCF} http://www.pwscf.org, 
S. Baroni {\it et al.},
Rev. Mod. Phys. 73, 515-562 (2001)

\bibitem{PBE} J.P.Perdew, K.Burke, M.Ernzerhof, 
Phys. Rev. Lett. {\bf 77}, 3865 (1996)

\bibitem{Calandra2005}  M. Calandra and F. Mauri, 
Phys. Rev. Lett. {\bf 95}, 237002 (2005), M. Calandra and F. Mauri,
Phys. Rev. B 74, 094507 (2006) 

\bibitem{footnote} To obtain the matrix ${\bm \Lambda}$
in eq. \ref{eq:Lambda} we 
consider a $3\times 3$ reduced matrix obtained summing over the contribution of
the two C nearest neighbours of the B atom (C') and the other 3 C atoms (C'') .


\bibitem{mcmillan} McMillan, 
Phys. Rev. {\bf 167}, 331 (1968).

\bibitem{Kong} Y. Kong {\it et al.},  
Phys. Rev. B {\bf 64}, 020501(R) (2001).

\bibitem{Multiband} A. Liu et al., Phys. Rev. Lett. {\bf } 87, 087005 (2001),
H.J. Choi et al., Phys.Rev. B {\bf 66}, 020513 (2002),
I.I. Mazin et al., Phys.Rev. B {\bf 69}, 056501 (2004),
A. Floris et al., Phys.Rev.Lett. {\bf 94}, 037004 (2005)

\bibitem{Sacepe} B. Sac\'ep\'e {\it et al.}, 
Phys. Rev. Lett. {\bf 96}, 097006 (2006) 


\end{thebibliography}
\end{document}